\documentstyle[12pt]{article}
\newlength{\dinwidth}
\newlength{\dinmargin}
\newcommand{\ba}{\begin{array}}
\newcommand{\ea}{\end{array}}
\newcommand{\be}{\begin{equation}}
\newcommand{\ee}{\end{equation}}
\newcommand{\bea}{\begin{eqnarray}}
\newcommand{\eea}{\end{eqnarray}}

\font\cmss = cmss12

\def\integer{{\rlap{\cmss Z} \hskip 1.8pt \hbox{\cmss Z}}}
\def\laplace{{\kern1pt\vbox{\hrule height 1.2pt\hbox{\vrule width 1.2pt\hskip
  3pt\vbox{\vskip 6pt}\hskip 3pt\vrule width 0.6pt}\hrule height 0.6pt}
  \kern1pt}}
\def\scriptlap{{\kern1pt\vbox{\hrule height 0.8pt\hbox{\vrule width 0.8pt
  \hskip2pt\vbox{\vskip 4pt}\hskip 2pt\vrule width 0.4pt}\hrule height 0.4pt}
  \kern1pt}}

\def\roughly#1{\raise.3ex\hbox{$#1$\kern-.75em\lower1ex\hbox{$\sim$}}}

\def\bee{\begin{eqnarray}}
\def\eee{\end{eqnarray}}
\def\be{\begin{equation}}
\def\ee{\end{equation}}

\begin{document}
\thispagestyle{empty}
\addtocounter{page}{-1}
\begin{flushright}
IASSNS-HEP 97/50\\
SNUTP 97-071\\
{\tt hep-th/9710245}\\
\end{flushright}
\vspace*{1.3cm}
\centerline{\Large \bf M(atrix) Theory on ${\bf T}_9/{\bf Z}_2$ 
Orbifold and Twisted Zero-Branes
\footnote{
Work supported in part by the NSF-KOSEF Bilateral Grant,
KOSEF SRC-Program, Ministry of Education Grant BSRI 97-2410, the 
Monell Foundation and the Seoam Foundation Fellowships, and the
Korea Foundation for Advanced Studies.}}
\vspace*{1.7cm} \centerline{\large\bf Nakwoo Kim${}^a$ and
Soo-Jong Rey${}^{a,b}$}
\vspace*{1cm}
\centerline{\large\it Physics Department, Seoul National University,
Seoul 151-742 KOREA${}^a$}
\vskip0.3cm
\centerline{\large\it School of Natural Sciences, Institute for
Advanced Study, Princeton NJ 08540 USA${}^b$}
\vspace*{1.5cm}
\centerline{\Large\bf Abstract}
\vskip0.4cm
M(atrix) theory compactified on an orbifold ${\bf T}_9/{\bf Z}_2$ is
studied. Via zero-brane parton scattering we find that each of the
$2^9 = 512$ orbifold fixed points carry $-1/32$ units of zero-brane
charge. The anomalous flux is cancelled by introducing a twisted sector
consisting of 32 zero-branes that are spacetime supersymmetry singlets. 
These twisted sector zero-branes are nothing but gravitational waves 
propagating along the M-theory direction. There is no D0-partons in the 
untwisted sector, a fact consistent with holographic principle. 
For low-energy excitations, the orbifold compactification is described by 
ten-dimensional supersymmetric Yang-Mills theory with gauge group $SO(32)$.
\vspace*{1.1cm}


\newpage

\section{Introduction}
\setcounter{equation}{0}
Via D0-partons, M(atrix) theory~\cite{bfss} defines a non-perturbative 
light-front Hamiltonian dynamics of M-theory~\cite{witten}, which unifies all 
known perturbative superstring theories at strong coupling regime.
Regularizing zero-momentum by compactifying the longitudinal M-direction on
a circle of radius $R$, the M(atrix) theory is defined by dynamics of 
$N$ D0-partons:
\cite{bfss} :
\be
S_M = {\rm Tr}_N \int \! d\tau \, \Big(
{1 \over 2 R} (D_\tau {\bf X}^I)^2 + {R \over 4} [{\bf X}^I , {\bf X}^J]^2
+ {\bf \Theta}^T D_\tau {\bf \Theta} + i R {\bf \Theta}^T \Gamma_I [{\bf X}^I ,
{\bf \Theta}] \Big).
\label{action}
\ee
where ${\bf X}^I$ and ${\bf \Theta}^\alpha$ denote 9 bosonic and 16 spinor
coordinates of 0-brane partons ($I = 1, \cdots, 9$ and $\alpha = 1, \cdots,
16$)~\footnote{
Our spinor conventions are as follows. We take Majorana representation so 
that $(16 \times 16)$ $\Gamma_I$'s are real and symmetric, 
$i {\overline {\bf \Theta}} \Gamma_- \equiv {\bf \Theta}^T$ and
$\Gamma^{(11)} {\bf \Theta} = + {\bf \Theta}$:
\be
\Gamma_i = \left( \begin{array}{cc}
0 & \sigma^i_{a {\dot a}} \\ \sigma^i_{{\dot a} a} & 0 \end{array} \right)
\hskip0.5cm i = 1, \cdots, 8; \hskip1cm
\Gamma_9 = \left( \begin{array}{cc}
- \delta_{a b} & 0 \\ 0 & + \delta_{{\dot a} {\dot b}} \end{array} \right).
\nonumber \\
\ee
}
The non-dynamical gauge field $A_\tau$ that enters through covariant
derivatives $D_\tau {\bf X}^I \equiv \partial_\tau {\bf X}^I -
i [A_\tau, {\bf X}^I]$ and
$D_\tau {\bf  \Theta}^\alpha \equiv \partial_\tau {\bf \Theta}^\alpha -
i [A_\tau, {\bf \Theta}^\alpha]$ projects the physical Hilbert space to
a gauge singlet sector and ensures invariance under area-preserving
diffeomorphism transformation. 
In the infinite-momentum light-front frame, out of thirty-two supersymmetries
of M-theory, only sixteen are realized as dynamical supersymmetries.
The other sixteen become kinematical supersymmetries.  
Thus, defining their supergenerators as $i \epsilon$ and $\xi$ respectively,
the M(atrix) theory is invariant under the following supersymmetry
transformations
\bee
\delta {\bf X}^I &=& - 2 \epsilon^T \Gamma^I {\bf \Theta} \nonumber \\
\delta {\bf \Theta} &=& { i \over 2} \Big( \Gamma_I D_\tau {\bf X}^I
+ {1 \over 2} \Gamma_{IJ} [{\bf X}^I, {\bf X}^J] \Big) \, \epsilon + \xi
\nonumber \\
\delta A_\tau &=& - 2 \epsilon^T \, {\bf \Theta}.
\label{susytransformation}
\eee
The sixteen dynamical and sixteen kinematical supersymmetry charges
are given by:
\bee
{\bf Q}_\alpha &=& {\sqrt R} {\rm Tr} \Big( \Gamma^I {\bf \Pi}_I +
{i  \over 2} \Gamma_{IJ} [{\bf X}^I, {\bf X}^J] \Big)_{\alpha \beta}
{\bf \Theta}_\beta,
\nonumber \\
{\bf S}_\alpha &=& {2 \over \sqrt R} {\rm Tr} {\bf \Theta}_\alpha
\label{susycharges}
\eee
respectively.

One outstanding issue in M(atrix) theory is proper description of M-theory
compactification. Generically, compactification onto shrinking $d$-dimensional 
space is defined in M(atrix) 
theory by a $(d+1)$-dimensional quantum theory with
an appropriate number of supersymmetries dictated by the holonomy of the space.
For $d \le 3$, the quantum theory is Yang-Mills gauge theory, which is 
(super)-renormalizable. For $d > 3$, the gauge theory becomes 
non-renormalizable, hence, should be replaced by yet-to-be-found fixed-point 
theory. Nevertheless, at low-energy and at appropriate limits of moduli space, 
the gauge theory should provide an effective M(atrix) field theory description 
to the compactification. 
As such, though admittedly limited, M(atrix) gauge theory description 
of compactification in {\sl all} dimensions should reveal many useful 
information of (compactified) quantum M-theory. 

Indeed, with such an effective M(atrix) theory approach, we have studied 
previously the ${\bf T}^5/\integer_2$ orbifold compactification
and were able to  many nontrivial M-theory dynamics inherent to this 
compactification including anomalous G-flux~\cite{mukhi, witten2, witten3} 
and spacetime spectrum~\cite{kimrey2}. 

In this paper, we keep this attitude, and study effective gauge theory 
description of M(atrix) theory compactified on ${\bf T}^9/\integer_2$ orbifold,
viz. compactification of all transverse dimensions. On ${\bf T}^9/\integer_2$ 
orbifold, $2^9 = 512$ fixed points are present and act as potential sources 
of anomalous charge. In section 2, we probe each orbifold fixed point via 
D0-parton scattering and find that it carries an anomalous {\sl gravity flux} 
of $ - 1/32$ unit. We find that cancellation of total {\sl gravity flux} then 
require introduction of a twisted sector consisting of 16 units of 
longitudinal gravitons and their $\integer_2$ images , viz. 32 twisted 
D0-branes. 
The resulting effective M(atrix) theory is thus given by ten-dimensional
supersymmetric Yang-Mills theory with gauge group SO(32), the only possible 
anomaly-free gauge theory in ten-dimensions.
In section 4, to understand the peculiarity of the M(atrix) theory on
${\bf T}^9/\integer_2$, we compare it with other previously studied 
orbifolds ${\bf S}^1/\integer_2$ and ${\bf T}^5/\integer_2$ with particular
attention to the realization of spacetime symmetry of M-theory as R-symmetry
of M(atrix) theory. We find, in ${\bf T}^9/\integer_2$ case, 
that the kinematical supersymmetry is completely projected out and  
that the twisted D0-branes are singlets under spacetime supersymmetry
even though they are not in M(atrix) theory parameter space.

\section{Anomalous Gravity Flux and Twisted Sector}
Consider ${\bf T}^9/\integer_2$ orbifold on which the M(atrix) theory is
compactified. On the orbifold there are $2^9 = 512$ fixed points. Near each 
orbifold fixed point, the D0-parton perceives the space locally as 
${\bf R}^9/\integer_2$. This orbifold is described by $\integer_2$ involution 
of M(atrix) theory defined on the covering space ${\bf R}^9$, viz. the
defining M(atrix) quantum mechanics Eq.~(1).
The Chan-Paton condition to the $\integer_2$ orbit is then given by:
\bee
{\bf X}^I &=& \, - \, M \cdot {\bf X}^{IT} \cdot M^{-1}
\nonumber \\
{\bf \Theta}_\alpha &=& \Gamma_\perp M \cdot {\bf \Theta}^T_\alpha
\cdot M^{-1} \quad ,
\quad \quad 
\Gamma_\perp = \Gamma^1 \cdots \Gamma^9.
\nonumber
\eee
Since $\Gamma_\perp = \Gamma^{(11)}$ in our convention and the spinor 
${\bf \Theta}$ is defined such that 
$\Gamma^{(11)} {\bf \Theta} = + {\bf \Theta}$, 
we find that the $\integer_2$ involution projects U(2N) covering space
gauge group to SO(2N) and remove the sixteen kinematical supersymmetries
completely. The latter peculiarity is rooted to the fact we have compactified 
{\sl all} transverse directions. In section 4, we will explain this 
peculiarity in yet another way in terms of M(atrix) theory R-symmetry and 
decomposition of kinematical and dynamical supersymmetries thereof. 

As in the other M(atrix) orbifolds previously studied 
${\bf S}_1/\integer_2$~\cite{kimrey1}, 
${\bf M}_2$ (M\"obius strip)~\cite{kimrey3},
${\bf T}^5/\integer_2$~\cite{kimrey2} and
${\bf T}^8/\integer_2$~\cite{ganor}, the fixed points of 
${\bf T}^9/\integer_2$ may carry anomalous M-theory charges and in turn
create uncancelled vacuum energy tadpole. In this section, using the
prescription defined in Ref.~\cite{kimrey1}, we explore the potential
anomalous M-theory charges via D0-parton scattering~\footnote{
After this part of
work was completed, we have learned an independent work by Ganor 
et.al.~\cite{ganor}, in which the same parton scattering result on 
${\bf T}^9/\integer_2$ as ours was obtained.}.

\subsection{Probing Gravity Flux via Parton Scattering}
We have previously prescribed fixed-target D0-parton 
scattering as a probe of potentially anomalous M-theory charges localized at 
orbifold fixed points~\cite{kimrey1}.
According to the prescription, the anomalous charge is probed by 
comparing the D0-parton scattering off the orbifold to that in the flat
space:
\be
{\cal V}_{\rm orientifold} (r, v) \equiv 
{\cal V}_{\integer_2 \quad \rm orbit}
(r, v) - {\cal V}_{\rm flat \quad space}
(r, v).
\ee

As a local probe, we place a 0-brane parton.
A 0-brane moving slowly near the fixed point will experience the presence
of fixed point or, equivalently, a mirror 0-brane parton approaching
toward the probing 0-brane parton. As such we expect that the effect
of orbifold fixed point amounts to a net force equivalent to
two-body 0-brane scattering. Dynamics of a zero-brane parton scattering
off the fixed point is described by SO(2) M(atrix) theory quantum
mechanics. Consider $A_9 = i vt \, \sigma_2/2$, where $v$ denotes
the {\sl relative} velocity between the probe zero-brane and the image
zero-brane. In the instantaneous limit, the two-body potential is
calculated from the forward scattering phase shift:

For the ${\bf T}_9/\integer_2$ orbifold we consider presently, we continue
utilizing the same prescription to probe possible anomalous charges. Since
the D0-parton scattering is a local process, we consider momentarily
region near any of the $2^9$ orbifold fixed points.

With the choice of $SO(2N)$ D0-parton scattering off the orbifold fixed point
is described by $SO(2)$ gauge theory. Thus, there is {\sl no} force coming
from the scattering. 
A straightforward quadratic expansion shows that there is {\sl no}
massive modes at all! This means that the probing 0-brane parton
moves freely in the orbifold space! This means that there is no
$v^4/r^7$ force one should expect from the consideration of
M(atrix) theory. We thus need to add a zero-brane to take into account
of the effects.

On the other hand, on the covering space, D0-D0 brane
scattering is described by $SU(2)$ quantum mechanics. Hence, we find that
the presence of $\Omega_9$ orientifold  
gives a net force:
\be
{\cal V}_{\rm orientifold} = 0 - {v^4 \over r^7} = - {v^4 \over r^7}.
\ee
Clearly, each orientifold carries negative anomalous D0-brane (anti-D0-brane) 
charges. On ${\bf T}_9/\integer_2$, D0-brane charnge conservation requires
introduction of twisted sector, which will maintain conservation of D0-brane
charge on ${\bf T}_9/\integer_2$. 
Consider a {\sl twisted}  
D0-brane of charge $Q$ located at the orientifold fixed point. By scattering
off D0-brane parton, we find that the twisted D0-brane exerts a force to
the parton of 
\be
{\cal V}_{\rm twisted} = 2 \cdot (2 Q) {(v/2)^4 \over (r/2)^7} 
= 2^5  Q {v^4 \over r^7}.
\ee
Here, the first factor of 2 is the difference of reduced mass between
the D0-parton and its image scattering versus D0-parton and twisted D0-brane
scattering and the second factor of 2 is the D0-brane charge as measured in
the covering space. 
We conclude that we need $- {1 \over 32}$ units of twisted D0-brane charge
localized at the orientifold fixed points. Had we taken USp gauge group, then
there is {\sl no} charge localized at the fixed point. 
This indicates that SO gauge group is the correct choice to the M(atrix)
theory. 

We thus find that local cancellation of anomalous gravity flux requires
introduction of a twisted sector consisting of thirty-two D0-partons.
One peculiarity of these twisted D0-partons as the spectrum of the 
twisted sector is that they are supersymmetric in M(atrix) theory side
while the corresponding twisted sector in two-dimensional M-theory 
are supersymmetry singlets~\cite{mukhi, witten2}. 
However, this is not a contradiction. It is well-known that purely bosonic
or fermionic adjoint multiplets that are singlets under supersymmetry
can exist in (0+1)-dimensions. Using this fact, we can associate the twisted 
sector D0-parton states with twisted sector spinors of (1+1)-dimensional
M-theory localized at each fixed points. These spinors are are chiral since 
they carry a positive BPS charge. Related aspects has been observed also 
in Ref.~\cite{mukhi3} with string and M-theory contexts.

What about total D0-parton flux conservation?
Since the transverse directions are all compactified (i.e. {\sl finite-in-all-directions} compactification), the total D0-parton flux has to vanish. 
The flux cannot leak to longitudinal direction since they are already 
boosted to the speed of light! 
These twisted D0-branes themselves. Furthermore, since D0-brane charge 
conservation is already saturated by the twisted sector, there is {\sl no} 
room for the D0-partons in the untwisted sector!

\subsection{Effective M(atrix) Gauge Theory on ${\bf T}^9/\integer_2$}
Let us now extend the analysis to ${\bf T}^9/\integer_2$ orbifold in the
limit the orbifold shrinks to zero size.
M(atrix) theory in this limit is conveniently described 
by first  T-dualizing the covering space ${\bf T}^9$ and then project 
onto $\integer_2$ orbits. 

The T-duality of M(atrix) theory is defined in terms of T-duality
transforamtion of D0-parton themselves. In the limit each sides of
${\bf T}^9/\integer_2$ orbifold shrinks to zero, infinitely many
multiply winding open string become light. They are more conveniently
describedonce all directions of covering space ${\bf T}^9$ is T-dualized so
that winding configurations are mapped into momentum modes.
Sum over all possible multiple winding quantum number is then mapped into
sum over all possible momentum quantum number, hence, Fourier transformation
of M(atrix) theory variables. The Fourier transformation results in 
ten-dimensional supersymmetric Yang-Mills theory on dual parameter space 
$\widetilde{\bf T}^9$. At least at classical level, the Yang-Mills  
theory is consistent with U(2N) covering space gauge group. 

The effective M(atrix)  gauge theory for ${\bf T}^9/\integer_2$ orbifold 
is then obtained by modding out with order-2 involution 
$\integer_2 = {\cal \Omega} \cdot {\cal P}$ of orientation reversal 
\bee
\Omega \quad : \quad A^M ({\bf y}, t) 
&\rightarrow& \Omega \cdot A^M(\Omega {\bf y}, t) \cdot \Omega^{-1}
= M \cdot A^{M \rm T} ({\bf y}, t) \cdot M^{-1}
\nonumber \\
\Psi_\alpha ({\bf y}, t)
&\rightarrow& \Omega \cdot \, \Psi_\alpha (\Omega {\bf y}, t) \, \cdot \Omega^{-1}
= M \cdot {\Psi_\alpha}^{\rm T}({\bf y}, t) \cdot M^{-1}.
\eee
and parity transformation
\bee
{\cal P} \quad : \quad
A^M ({\bf y}, t) 
&\rightarrow& {\cal P} \cdot A^M({\cal P} {\bf y}, t) \cdot {\cal P}^{-1}
= \, - \, A^M( +{\bf y}, t)
\nonumber \\
\Psi_\alpha
({\bf y}, t) &\rightarrow& {\cal P} \cdot \Psi_\alpha ({\cal P} {\bf y}, t)
\cdot {\cal P}^{-1} = \Gamma^{(11)} \Psi_\alpha (+{\bf y}, t).
\eee
Noting that $\Gamma^{(11)} \Psi = + \Psi$ in our convention, 
we find, at least classically, the effective M(atrix) gauge theory 
for ${\bf T}^9/\integer_2$ orbifold compactification is given by 
ten-dimensional supersymmetric gauge theory with gauge groupi SO(2N) or any 
sub-group thereof.

We have already identified the twisted sector consists of sixteen D0-branes 
and their images. On ${\bf T}^9/\integer_2$, they are localized at 512 
orbifold fixed points.
After T-duality on all directions, however, these twisted sector D0-branes
are delocalized and fill in the bulk of dual parameter space 
$\widetilde{\bf T}^9$. We thus conclude that the twisted sector is also 
described by ten-dimensional Yang-Mills theory! In fact, in ten-dimensional
${\cal N} = 1$ rigid supersymmetry, there is no other possible 
supermultiplets than Yang-Mills one.
Thus it is not surprising that both untwisted and twisted sector are 
described by Yang-Mills gauge theory.

Combining both the untwisted and the twisted sectors, the effective M(atrix) 
gauge theory follows from Fourier transform of $\integer_2$ invariant 
configuration of $2(N + 16)$ D0-partons and the result is ten-dimensional
supersymmetric Yang-Mills theory with gauge group SO(2N + 32).
Quantum consistency of the M(atrix) gauge theory requires that the 
gauge group is free from gauge anomalies. This singles out the gauge 
group to be SO(32), hence, $N=0$! This effective M(atrix) gauge theory with 
{\sl vacuous untwisted sector} is precisely what agrees with the simple 
argument given earlier that on ${\bf T}^9/\integer_2$ (in fact for any
"{\sl finite-in-all-directions}" compactification) there is no transverse 
space left over into which the D0-parton flux can extend over.  
The only D0-partons one can place on ${\bf T}^9/\integer_2$ are the 32 
ones which neutralize the anomalous gravity charges localized at the
orbifold fixed points.
Dynamics of these neturalizing D0-partons are described by ten-dimensional 
SO(32) supersymmetric Yang-Mills gauge theory.
\section{Spacetime Symmetry of M-theory versus R-Symmetry of M(atrix) Theory}
In due course of studying effective M(atrix) gauge theory, we have 
encountered several peculiarities unique to ${\bf T}^9/\integer_2$ 
compactification. 
In particular we have seen that the untwisted sector D0-partons are completely
suppressed and that the twisted sector spectrum that are two-dimensional
spacetime M-theory supersymmetry singlets are described in M(atrix) gauge 
theory by Yang-Mills supermultiplets of ten-dimensional parameter space.
In order to have better understanding of these peculiarities, 
we now compare ${\bf S}^1/\integer_2$, ${\bf T}^5/\integer_2$
and ${\bf T}^9/\integer_2$ orbifold compactifications in a unified manner.
In particular, we pay attention to the way spacetime Lorentz symmetry in
M-theory is realized in terms of R-symmetry in corresponding 
M(atrix) gauge theory.

The little group of spacetime Lorentz symmetry of M-theory is SO(9). In 
M(atrix) theory description of M-theory, this symmetry is realized as 
R-symmetry of extended supersymmetry.
Under $SO(9)_R$ of M(atrix) theory, nine transverse space fields $X^i$'s
transform as vector representation, while fermionic superpartners transform as 
sixteen-dimensional spinor representation.
Let us see how the one-to-one correspondence between spacetime symmetry
and R-symmetry is realized upon compactification of M-theory (under which 
the M(atrxi) theory becomes decompactified).
In particular, we are interested in the realization of supercharges in 
both descriptions. 

Once compactified on ${\bf S}_1$, 
M-theory reduces to Type IIA superstring, whose supercharges are two 
Majorana-Weyl spinors of opposite (9+1)-dimensional chirality, hence, 
nonchiral. 
The corresponding M(atrix) theory is (1+1)-dimensional supersymmetric
Yang-Mills theory with sixteen supercharges. It turns out that the
M(atrix) theory is also nonchiral, consisting of eight (1+1)-dimensional 
Majorana-Weyl spinors for each chirality. 
The sixteen supercharges are thus realized as (8,8). 
Upon compactification, the R-symmetry is now reduced to $SO(8)_R \subset
SO(9)_R$. Under the R-symmetry, the eight right-moving supercharges 
transform as spinor representation while the eight left-moving supercharges
transform as conjugate spinor representation. 
It is also obvious that half of the M-theory supercharges that become one
right-handed d=10 Majorana-Weyl spinor become eight M(atrix) theory 
supercharges that are right-handed d=2 Majorana-Weyl spinors. 
We thus find that spacetime symmetry of M-theory is realized as R-symmetry
of M(atrix) theory, in which the spinor chirality in each sides are 
correlated via spinor decomposition. 

Heterotic M(atrix) theory is obtained by taking a quotient of $\integer_2$
involution. The involution projects out supercharges in left-handed spinor
representation in both M-theory and M(atrix) theory sides. It should then
be noted that the {\sl twisted sectors} in each side have left-handed 
chirality while supercharges are right-handed. 
In M-theory side, the twisted sector consists of d=10, (1,0) 
 gauge supermultiplet. In M(atrix) theory side, the twisted sector
fields are d=2, (8,0) supersymmetry singlet, left-handed spinors.

As a next nontrivial compactification, consider M-theory compactified on 
${\bf T}_5$. M-theory on ${\bf T}_5$ has six-dimensional (2,2) supersymmetry
consisting of two Weyl supercharges of each chirality.
The original spacetime symmetry SO(10,1) is reduced to SO(5,1)$\times$SO(5).
the latter is nothing but R-symmetry in d=6 M-theory. With two Weyl spinors
in d=6, we have global symmetry USp(4) = SO(5).
As before, the little group of M-theory spacetime symmetry
is SO(4)$\subset$SO(5,1). The corresponding M(atrix) theory is 
d=6 super-Yang-Mills theory whose supercharges are (1,1). 
In M(atrix) theory side, the little group is realized as R-symmetry, 
viz., SO(4) $\subset$ SO(9). 
The sixteen supercharges are realized as two d=6 Weyl spinors of both
chirality. Each Weyl spinors give rise to USp(2) = SU(2) global symmetry, 
which are one part of SO(4) = SU(2)$\times$SU(2). The M-theory supercharges
, which were chiral, are now nontrivial for only one of the two SU(2)'s
in M(atrix) theory. Again, M-theory and M(atrix) theory chiralities of 
supercharges are related each other via decomposition of spinors. 
Upon $\integer_2$ quotient, we drop left-handed Weyl supercharge in M-theory
and corresponding R-charged supercharge in M(atrix) theory, which is again
left-handed. The twisted sectors in each descriptions are hypermultiplets
and fermions all transform as right-handed. 

Let us proceed further and consider M-theory compactification on ${\bf T}_9$. 
In M-theory side, we have D=2, (16,16) supersymemtry, viz, supercharges 
comprise of sixteen right-handed and sixteen left-handed Majorana-Weyl
spinors. 
The M-theory has SO(9) R-symmetry and each sixteen supercharges transform as 
Majorana spinor representation of SO(9) R-symmetry.
The corresponding M(atrix) theory is d=10 super-Yang-Mills theory. The 
M(atrix) theory supercharge consist of single, sixteen dimensional 
Majorana-Weyl spinor.
An amusing point is that the M-theory is nonchiral but the M(atrix) theory
is chiral. This can be traced back to the fact we have chosen D0-branes,
not anti-D0-branes, as M(atrix) theory partons.
Therefore, in this case, the other supercharge of opposite chirality is
hidden as kinematical supersymmetry charge. Upon $\integer_2$ orientifolding,
we drop left-handed supercharges in M-theory and kinematical supercharge
in M(atrix) theory.

\section{Discussions}
In this paper we have studied M(atrix) theory description of
compactified M theory on ${\bf T}^9/\integer_2$ orbifold. In the large
volume limit, through zero-brane parton scattering off the orbifold
fixed points, we have found that the fixed point carries anomalous
gravi-photon flux of one unit. In order to cancel the anomalous flux
a twisted sector consisting of thirty-two D0-partons is introduced. 
Effectively, the 512 states of the D0-parton BPS multiplet that are 
split due to spontaneously broken kinematical supersymmetry and cancels
locally the anomalous gravity flux at each of $512$ orbifold fixed points. 

Because all the nine transverse directions are compactified, we have 
seen that the ${\bf T}^9/\integer_2$ orbifold entails certain peculiarities 
not encountered in lower dimensional compactifications. 
In the infinite momentum frame, the D0-parton flux 
are extended entirely into transverse directions.
Since all transverse directions are compactified, the D0-partons charge
conservation requires that no D0-partons to be present. This {\sl thinning}
of parton degrees of freedom, however, stops at 
thirty-two for ${\bf T}^9\integer_2$, since the orbifold fixed points
carry anomalous gravity flux and thirty-two D0-partons are needed to cancel
them. 
We have shown that this fits nicely with the fact that effective M(atrix)
gauge theory is the unique anomaly-free SO(32) supersymmetric Yang-Mills
theory in ten dimensions.

\vskip0.5cm
S.-J.R. thanks to D. Kabat, S. Sethi and E. Witten for helpful 
discussions and the Institute for Advanced Study for hospitality, 
where part of this work was done.


\end{document}